\documentclass[twocolumn,preprintnumbers,amsmath,amssymb,floatfix]{revtex4}

\usepackage{graphicx}% Include figure files
\usepackage{dcolumn}% Align table columns on decimal point
\usepackage{bm}% bold math

\begin{document}

\title{Centrality dependence of charged hadron transverse momentum spectra  \\ in Au+Au collisions from $\sqrt{s_{_{\it NN}}} =$~62.4 to 200 GeV}

% updated author list on 29 April 2004
\author{
% Authors for data EXCLUSIVELY from RUN2004 which included AuAu @ 63 & 200 GeV
% and pp @ 200 GeV
%
% Last edited 9-Apr-2004 by George Stephans\\  \vspace{0.2in}
%
B.B.Back$^1$,
M.D.Baker$^2$,
M.Ballintijn$^4$,
D.S.Barton$^2$,
R.R.Betts$^6$,
A.A.Bickley$^7$,
R.Bindel$^7$,
W.Busza$^4$,
A.Carroll$^2$,
Z.Chai$^2$,
M.P.Decowski$^4$,
E.Garc\'{\i}a$^6$,
T.Gburek$^3$,
N.George$^2$,
K.Gulbrandsen$^4$,
C.Halliwell$^6$,
J.Hamblen$^8$,
M.Hauer$^2$,
C.Henderson$^4$,
D.J.Hofman$^6$,
R.S.Hollis$^6$,
R.Ho\l y\'{n}ski$^3$,
B.Holzman$^2$,
A.Iordanova$^6$,
E.Johnson$^8$,
J.L.Kane$^4$,
N.Khan$^8$,
P.Kulinich$^4$,
C.M.Kuo$^5$,
W.T.Lin$^5$,
S.Manly$^8$,
A.C.Mignerey$^7$,
R.Nouicer$^{2,6}$,
A.Olszewski$^3$,
R.Pak$^2$,
C.Reed$^4$,
C.Roland$^4$,
G.Roland$^4$,
J.Sagerer$^6$,
H.Seals$^2$,
I.Sedykh$^2$,
C.E.Smith$^6$,
M.A.Stankiewicz$^2$,
P.Steinberg$^2$,
G.S.F.Stephans$^4$,
A.Sukhanov$^2$,
M.B.Tonjes$^7$,
A.Trzupek$^3$,
C.Vale$^4$,
G.J.van~Nieuwenhuizen$^4$,
S.S.Vaurynovich$^4$,
R.Verdier$^4$,
G.I.Veres$^4$,
E.Wenger$^4$,
F.L.H.Wolfs$^8$,
B.Wosiek$^3$,
K.Wo\'{z}niak$^3$,
B.Wys\l ouch$^4$\\
\vspace{3mm}
\small
%
% Note that this is the full form of the addresses, for conference proceedings,
% you can use the reduced one that follows
%
% $^1$~Physics Division, Argonne National Laboratory, Argonne, IL 60439-4843,
% USA\\
% $^2$~Chemistry and C-A Departments, Brookhaven National Laboratory, Upton, NY
% 11973-5000, USA\\
% $^3$~Institute of Nuclear Physics, Krak\'{o}w, Poland\\
% $^4$~Laboratory for Nuclear Science, Massachusetts Institute of Technology,
% Cambridge, MA 02139-4307, USA\\
% $^5$~Department of Physics, National Central University, Chung-Li, Taiwan\\
% $^6$~Department of Physics, University of Illinois at Chicago, Chicago, IL
% 60607-7059, USA\\
% $^7$~Department of Chemistry, University of Maryland, College Park, MD 20742,
% USA\\
% $^8$~Department of Physics and Astronomy, University of Rochester, Rochester,
% NY 14627, USA\\
%
%
$^1$~Argonne National Laboratory, Argonne, IL 60439-4843, USA\\
$^2$~Brookhaven National Laboratory, Upton, NY 11973-5000, USA\\
$^3$~Institute of Nuclear Physics PAN, Krak\'{o}w, Poland\\
$^4$~Massachusetts Institute of Technology, Cambridge, MA 02139-4307, USA\\
$^5$~National Central University, Chung-Li, Taiwan\\
$^6$~University of Illinois at Chicago, Chicago, IL 60607-7059, USA\\
$^7$~University of Maryland, College Park, MD 20742, USA\\
$^8$~University of Rochester, Rochester, NY 14627, USA\\
}

\begin{abstract}\noindent

We have measured transverse momentum distributions of charged hadrons
produced in Au+Au collisions at $\sqrt{s_{_{\it NN}}} =$ 62.4 GeV.
The spectra are presented for transverse momenta $0.25 < p_T < 4.5$~GeV/c,
in a pseudo-rapidity range of $0.2 < \eta < 1.4$.
The nuclear modification factor $R_{AA}$ is calculated relative to p+p
data at the same collision energy as a function of collision centrality.
For $p_T > 2$~GeV/c, $R_{AA}$ is found to be significantly larger than in
Au+Au collisions at $\sqrt{s_{_{\it NN}}} =$ 130 and 200 GeV. In
contrast, we find that the evolution of the invariant yields per 
participant pair
from peripheral to central collisions is approximately 
energy-independent  over this range of
collision energies. This observation challenges models of
high $p_T$ hadron suppression in terms of parton energy loss.

\vspace{3mm}
\end{abstract}

\maketitle

The yield of charged hadrons produced in collisions of gold nuclei at 
an energy of $\sqrt{s_{_{\it NN}}} = 62.4$~GeV has been measured 
with the PHOBOS detector at the Relativistic Heavy Ion Collider (RHIC) at Brookhaven National Laboratory.
The data are  presented as a function of transverse momentum ($p_T$) and collision centrality.
% problem sentence ->
The goal of these measurements is to study the modification of particle production in the presence of the produced medium by comparing to nucleon-nucleon collisions at the same energy.

This measurement was motivated by  results from Au+Au collisions at $\sqrt{s_{_{\it NN}}} =$~130 and 200~GeV.
Hadron production at these energies was found to be strongly suppressed
relative to expectations based on an independent superposition of  nucleon-nucleon collisions at 
$p_T$ of 2--10 GeV/c \cite{phenix_quench,phenix_highpt_npart,star_highpt_npart, phobos_highpt_npart}. 
The modification of high-$p_T$ hadron yields has commonly been investigated using the nuclear modification 
factor $R_{\it AA}$, defined as 
\begin{equation}
R_{\it AA} = \frac{\sigma_{pp}^{inel}}{\langle N_{\it coll} \rangle} 
              \frac{d^2 N_{\it AA}/dp_T d\eta} {d^2 \sigma_{pp}/dp_T d\eta}.
\end{equation}
A value of $R_{\it AA} = 1$ corresponds to scaling of particle production with the average number of 
binary nucleon-nucleon collisions, $\langle N_{coll} \rangle$, within a heavy-ion collision. 

For the production of charged hadrons in 
central Au+Au collisions at $\sqrt{s_{_{\it NN}}} =$~200~GeV,
values of $R_{\it AA} \approx  0.2$ are observed at $p_{T}=4$~GeV/c \cite{phenix_highpt_npart,star_highpt_npart,phobos_highpt_npart}.
Such a suppression had been predicted to occur as a consequence of the energy loss of high-$p_T$ partons
in the dense medium formed in Au+Au collisions \cite{jet_quench_theory}.
This hypothesis is consistent with the observed absence of this effect in deuteron--gold collisions at the same collision
energy \cite{phenix_dAu,star_dAu,brahms_dAu,phobos_dAu}.

% remove 'extensive'?
While studies of the system-size dependence of the high-$p_T$ suppression exist,
the dependence on the collision energy 
is poorly known. Data on the production of neutral pions in Pb+Pb collisions 
at $\sqrt{s_{_{\it NN}}} =  17.2$~GeV suffer from large uncertainties in the 
 parametrizations of the p+p reference data \cite{wa98spectra,wang,dEnterria}. 
The results presented here for Au+Au collisions at $\sqrt{s_{_{\it NN}}} = 62.4$~GeV,
in combination with the results from 130 and 200 GeV collisions, 
allow the first detailed examination of the connection between high $p_T$ 
suppression and collision energy, charged-particle density and collision geometry.

The data were collected using the PHOBOS 
two-arm magnetic spectrometer. 
Details of the experimental setup can be found in \cite{phobos_nim}.
The primary event trigger used the time difference between signals in 
two sets of 10 \v Cerenkov counters, located at $4.4< | \eta | <4.9$, to select collisions along the beam-axis that were close to the nominal vertex position.

\begin{figure}[t]
\hspace{-1cm}
\includegraphics[height=8.1cm]{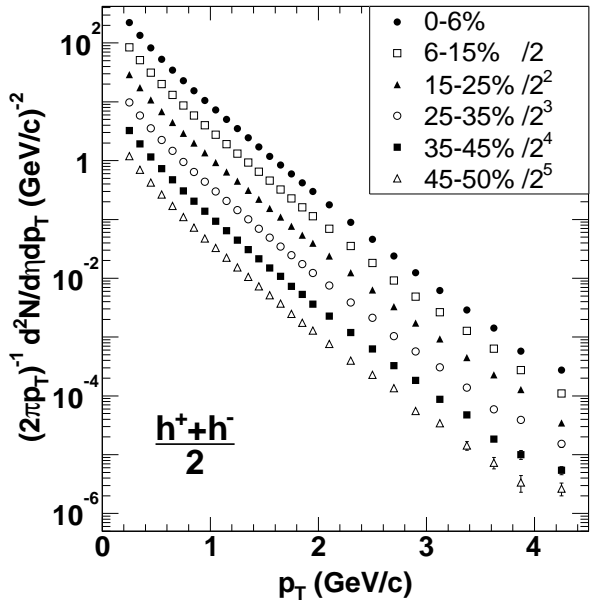}
\caption{ \label{SpectraCentBins}
Invariant yields for charged hadrons from Au+Au collisions at $\sqrt{s_{_{\it NN}}} = $ 62.4~GeV, in the pseudo-rapidity interval $0.2 < \eta < 1.4$ as a function of $p_T$ for 6
centrality bins. For clarity, consecutive bins are scaled by factors of 2.
The systematic uncertainties are smaller than the symbol size.}
\end{figure}

For the analysis presented here, events were divided into six centrality classes,
based on the observed signal in two sets of 16 scintillator 
counters covering pseudo-rapidities 
$3.0 < |\eta |< 4.5$. 
The results of the Glauber calculation implemented in the HIJING model
\cite{phobos_cent_200,hijing} 
were used
to estimate the average number of
participating nucleons, $\langle N_{part} \rangle$, and binary collisions, $\langle N_{coll} \rangle$, for each
centrality class. For the Glauber calculation, as well as for the determination of $R_{\it AA}$ at 62.4~GeV, we used
$\sigma_{pp}^{inel} = 36 \pm 1$~mb. 
The values and systematic uncertainties for $\langle N_{part} \rangle$ 
and $\langle N_{coll} \rangle$ at both 62.4 and 200~GeV are listed in Table~I.

\begin{table}
\caption{
\label{table1}
Details of the centrality classes used in this analysis.  
Bins are expressed in terms of percentage of the total inelastic Au+Au cross-section.}
\begin{ruledtabular}
\begin{tabular}{lllll}
Centrality &
$ \left<N_{part}^{62.4}\right>$ &  $ \left< N_{coll}^{62.4}\right>$ 
& $ \left<N_{part}^{200}\right>$ &  $ \left< N_{coll}^{200}\right>$ \\
\hline
45--50\% &  $61 \pm 7$  & $76   \pm 12$   &  $65 \pm 4$   & $107 \pm 16$   \\
35--45\% &  $86   \pm 9$  & $120  \pm 16$  &  $93  \pm 5$  & $175 \pm 25$  \\
25--35\% &  $130  \pm 10$ & $215  \pm 22$  &  $138 \pm 6$ & $300 \pm 39$  \\
15--25\% &  $189  \pm 9$  & $370  \pm 24$  &  $200 \pm 8$ & $500 \pm 60$  \\
6--15\%  &  $266  \pm 9$  & $590  \pm 21$  &  $276 \pm 9$ & $780 \pm 86$  \\
0--6\%   & $335  \pm 11$  & $820  \pm 25$  &  $344 \pm 11$ & $1050 \pm 105$ \\
%\hline
\end{tabular}
\end{ruledtabular}
\end{table}

The event selection and track reconstruction procedure for this analysis closely  follows 
the procedure for the previously published analysis at $\sqrt{s_{_{\it NN}}} = 200$~GeV \cite{phobos_highpt_npart}. 
Events with a primary vertex position within $\pm 10$~cm of
the nominal vertex position were selected.
Only particles traversing a full spectrometer arm were included in the analysis, resulting in 
a low transverse momentum cutoff at $p_T \approx 0.2$~GeV/c.

\begin{figure}[t]
\hspace{-0.6cm}
\includegraphics[height=7.0cm]{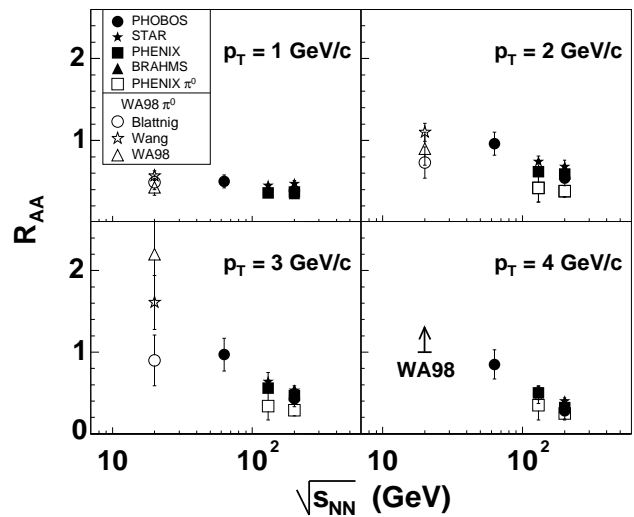}
\caption{ \label{RAAVsEnergy}
Nuclear modification factor $R_{\it AA}$ for central A+A events, as a function of collision energy for $p_T = $~1, 2, 3 and 4 GeV/c. 
Filled symbols show data for charged hadrons, open symbols show $\pi^0$ data.
For the WA98 $\pi^0$ data, $R_{\it AA}$ is shown using three different parameterizations for the p+p reference spectrum \cite{wa98spectra,wang,dEnterria}; for $p_T = 4$~GeV/c, the arrow indicates the lower limit on $R_{\it AA}$.
The error bars show the combined systematic and statistical uncertainty obtained by interpolating the $p_T$ dependence
of $R_{\it AA}$.}
\end{figure}

\begin{figure*}[t]
\includegraphics[height=10.5cm]{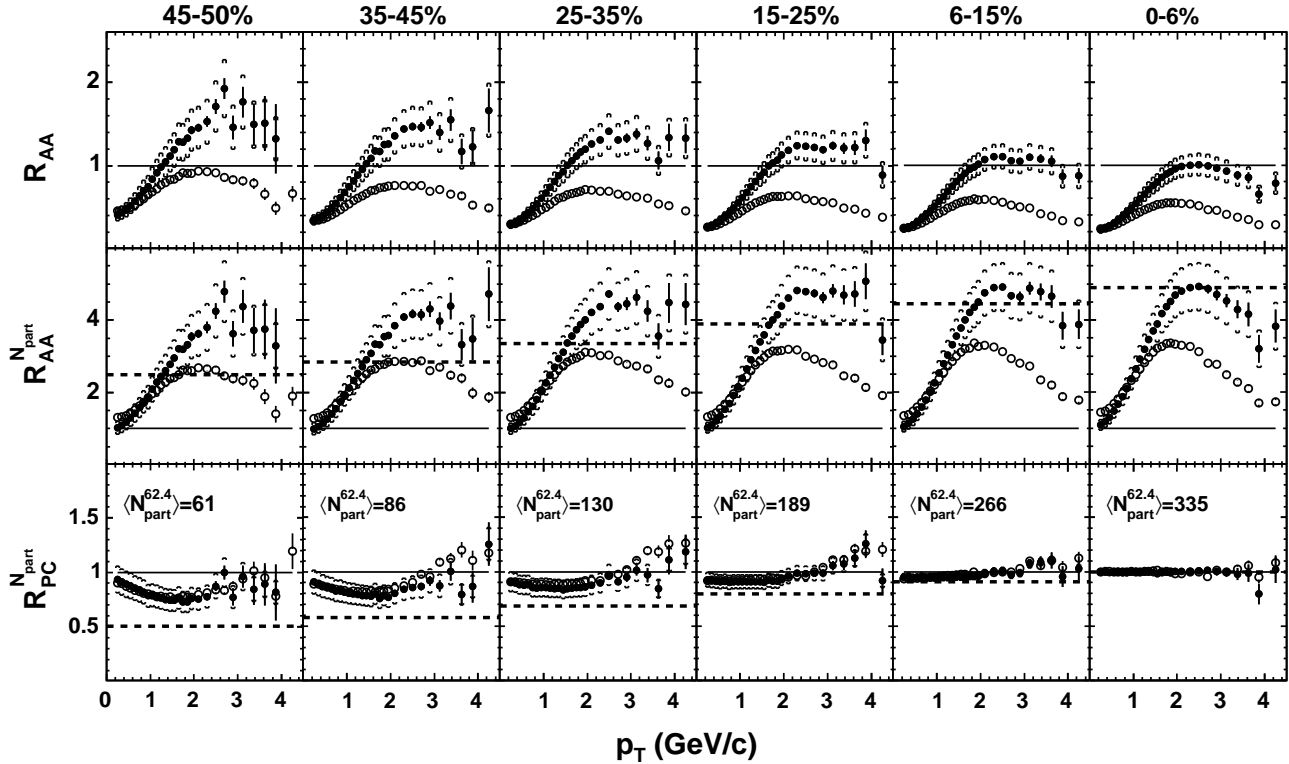}
\caption{ \label{MasterPlot}
Ratio of $p_T$ distributions from Au+Au collisions to various reference distributions 
at $\sqrt{s_{_{\it NN}}} = 62.4$~GeV (filled symbols) and 
200~GeV (open symbols). Data are shown in six bins of centrality, ranging 
from $\langle N_{part} \rangle = 61$ to 335 for 62.4 GeV collisions. The top row shows the nuclear 
modification factor $R_{\it AA}$, i.e.\ the ratio relative to proton-(anti)proton collisions scaled by $\langle N_{coll} \rangle$. 
The middle row shows $R_{\it AA}^{N_{part}}$, which uses proton-(anti)proton spectra scaled by 
$\langle N_{part}/2 \rangle$. The bottom row shows 
$R_{\it PC}^{N_{part}}$, using a fit to central data scaled by $\langle N_{part}/2 \rangle$ as a reference.
The dashed line in the middle and bottom rows indicates the expectation for $\langle N_{coll} \rangle$ scaling at $\sqrt{s_{_{\it NN}}} = 62.4$~GeV
relative to the reference distribution.
Systematic uncertainties for all plots are shown by brackets (90\% C.L.)}
\end{figure*}

The transverse momentum distribution for each centrality bin was corrected separately for the geometrical 
acceptance of the detector, the
inefficiency of the tracking algorithm, secondary and incorrectly 
reconstructed particles,  and the distortion due to binning and momentum
resolution. The relative importance of most of the corrections and the associated estimated systematic uncertainties 
are similar to those reported in the previous analysis \cite{phobos_highpt_npart}. The exception is the 
correction for momentum resolution and binning effects
at high $p_T$, where changes to the track fitting procedure improved the resolution 
and reduced the corresponding uncertainty by a factor of two. 

In Fig.~\ref{SpectraCentBins}, we present the invariant yield
of charged hadrons as a function of transverse momentum, obtained by
averaging the yields of positive and negative hadrons. 
Data are shown for six centrality bins and are averaged over a pseudo-rapidity interval $0.2 < \eta < 1.4$.

Focusing first on central events, Fig.~\ref{RAAVsEnergy} 
shows the nuclear modification factor $R_{\it AA}$ for the 6\% most central 
Au+Au collisions at $\sqrt{s_{_{\it NN}}} = 62.4$~GeV for four different values of $p_T$ ranging from 1 to 4 GeV/c.
% should these be pt bins, not values; what's their width? [Alice]
Results from p+p collisions at the same energy \cite{ISR} were used 
in the calculation of $R_{\it AA}$.
The new results are compared with results for charged hadrons at $\sqrt{s_{_{\it NN}}} =  130$ and 200 GeV. 
For $p_T = 2$~GeV/c and above,  we observe a smooth decrease
of $R_{\it AA}$ from 62.4 to 200 GeV in central Au+Au collisions. 

Also shown in Fig.~\ref{RAAVsEnergy} is $R_{AA}$ for $\pi^0$ production in Pb+Pb collisions at 17.2 GeV from WA98, 
obtained using three p+p reference parametrizations \cite{wa98spectra,wang,dEnterria}, as well
as $\pi^0$ data for 130~GeV and 200~GeV from PHENIX \cite{phenix_pi0_130,phenix_pi0_200}. 
Although the uncertainties in the 17.2 GeV reference distribution are large, 
the data indicate that $R_{AA}$ for $\pi^0$ production at $p_T > 3$~GeV/c 
drops from $R_{\it AA} > 1$ at $\sqrt{s_{_{\it NN}}} =  17.2$~GeV to $R_{\it AA} < 0.2$ at $\sqrt{s_{_{\it NN}}} = 200$~GeV.
It is interesting to note that the change in the 
observed charged hadron pseudo-rapidity density over the same energy range is only a factor of two \cite{phobos_lim_frag}.
The data also show that $R_{\it AA}$ for neutral pions is consistently lower than for charged hadrons at the same collision energy,
which is important when comparing data from
different experiments.

The centrality evolution of $R_{\it AA}$ for the 62.4~GeV data can be studied in detail in the top row of Fig.~\ref{MasterPlot}.
For comparison, we include the results from Au+Au collisions at 200 GeV \cite{phobos_highpt_npart},
using the same centrality binning. 
We observe that $R_{\it AA}$ reaches a maximum
of approximately 1.6 for peripheral collisions at 62.4~GeV, whereas the maximum value 
observed at 200 GeV is close to 1. For both datasets,
a significant and systematic decrease in $R_{\it AA}$ is observed in progressing from peripheral to central events.

It has been previously noted that the observed strong centrality dependence of $R_{\it AA}$ at $\sqrt{s_{_{\it NN}}} =$ 200~GeV
corresponds to a relatively small change in the yield per participating nucleon
\cite{phobos_highpt_npart}
% between peripheral and central Au+Au collisions  
and that over the same centrality range, the total yield of charged
particles per participating nucleon is constant within the experimental uncertainties \cite{universality}. 
These observations lead us to define 
\begin{equation}
R_{\it AA}^{N_{part}} = \frac{\sigma_{pp}^{inel}}{\langle N_{\it part}/2 \rangle} 
              \frac{d^2 N_{\it AA}/dp_T d\eta} {d^2 \sigma_{pp}/dp_T d\eta},
\end{equation}
in analogy to Equation~1, where now we scale the reference spectrum by $N_{part}/2$, rather than $N_{coll}$.
The centrality, $p_T$ and energy dependence of $R_{\it AA}^{N_{part}}$ is shown in the middle row of Fig.~\ref{MasterPlot}
for collisions at 62.4~GeV and 200~GeV.
Over the full range of $p_T$ studied here, the yield per participant pair shows a 
variation of only $\approx 25$\% from peripheral to central collisions for both collision energies.

This centrality independence is further illustrated in the bottom row of Fig.~\ref{MasterPlot}. The quantity $R_{\it PC}^{N_{part}}$, defined as 
\begin{equation}
R_{\it PC}^{N_{part}} = \frac{\langle N_{\it part}^{0-6\%} \rangle}{\langle N_{\it part} \rangle} 
              \frac{d^2 N_{\it AA}/dp_T d\eta} {d^2 N_{\it AA}^{0-6\%}/dp_T d\eta},
\end{equation}
is shown as a function of $p_T$ for the six centrality bins.
$R_{\it PC}^{N_{part}}$ measures the change in yield per participant pair,
relative to a fit to the central data. Normalizing to the central, rather than the peripheral, data
has the advantage that the results are more easily compared among different experiments with different ranges 
in centrality, while providing the same information in comparisons with theoretical calculations. 
Data are shown for collisions at 62.4~GeV and 200~GeV. This plot again shows the small variation
of the yield per participant pair from peripheral to central collisions. 
Furthermore, it demonstrates that the modification of the 
yield from peripheral to central collisions is the same for both energies, over the full $p_T$ range, within experimental 
uncertainties of less than $10\%$. This striking agreement can be compared with the much larger  variation of $R_{\it AA}$ 
as a function of energy, centrality and $p_T$.

This measurement of the centrality dependence of particle yields at $\sqrt{s_{_{\it NN}}} =$~62.4~GeV severely 
constrains the parameter space for any model
of high-$p_T$ hadron suppression in terms of parton energy loss. It remains to be seen whether the agreement in the centrality
evolution of the yields per participant pair, which ranges over a factor of three in collision energy, is a natural consequence of such models.
It is important to note that the energy-independence of the centrality evolution is a characteristic feature, not only of 
total and differential particle yields \cite{universality, Back:2002ft}, but also of multi-particle correlation measurements 
such as Bose-Einstein correlations \cite{hbt}. 
The apparent dominance of the initial-state geometry,
even for observables closely related to the dynamical evolution of heavy-ion collisions,
is one of the key features of these interactions that remains to be understood.

% New acknowledgements taken from phobos_acknowledge.tex
% Last edited 26-Feb-2004 by George Stephans
We acknowledge the generous support of the Collider-Accelerator Department at BNL for providing the 62.4~GeV beams.
This work was partially supported by US DoE grants DE-AC02-98CH10886,
DE-FG02-93ER40802, DE-FC02-94ER40818, DE-FG02-94ER40865,
DE-FG02-99ER41099, W-31-109-ENG-38, US NSF grants 9603486, 0072204 and 0245011, Polish KBN grant 2-P03B-10323, and NSC of Taiwan contract NSC
89-2112-M-008-024.

\end{document}